\documentstyle[aps,prd,multicol]{revtex}
\tighten

\begin{document}
\title{Dirac and Majorana heavy neutrinos at LEP II}
\author{F.M.L. Almeida Jr.,Y. A. 
Coutinho,
J. A. Martins Sim\~oes\thanks{E-mail: simoes@.if.ufrj.br}, 
M.A.B. do Vale and S. Wulck\\
Instituto de F\'\i sica\\
Universidade Federal do Rio de Janeiro, RJ, Brazil \\}
\maketitle
\begin{abstract}
\par
The possibility of detecting single heavy Dirac and/or Majorana neutrinos at LEP II is investigated  for heavy neutrino masses in the range $M_N=(\sqrt{s}/2,\sqrt{s})$. We study the process $e^+e^- \longrightarrow \nu \,e^\pm  \, q_i\, \bar q_j $ as a clear signature for heavy neutrinos. Numerical estimates for cross sections and distributions for the signal and the background are calculated and a Monte Carlo reconstruction of final state particles after hadronization is presented.\\
\par
\end{abstract}
\begin{multicols}{2}
\section{Introduction}\setcounter{equation}{0}
The recently accumulated data at LEP offers a unique opportunity for the search of new particles. At center of mass energies around 200 GeV and a combined integrated luminosity of nearly 2500 pb$^{-1}$ \cite{LEP} this experimental facility can be employed, for instance, in the search of new heavy Dirac and/or Majorana neutrinos. The L3 Collaboration \cite{L3a} at LEP has recently published results on the search of new possible excited charged and neutral leptons. Their experimental data was compared with a specific model of new excited leptons interacting with the Standard Model gauge bosons and new limits were found. Most of their data was compared with pair production of new heavy leptons. Single production of new possible charged leptons according to Standard Model extentions is also under study at the Delphi Collaboration \cite{ALM} at LEP. A natural question is then: what are the consequences of these  results for other models of new heavy leptons? This is particularly important for new heavy neutrinos. The recent SNO results \cite{SNO} provide increasing evidence for light neutrino oscillations and non zero neutrino masses. A possible explanation of the smallness of neutrino masses is the ``see-saw" mechanism, which implies new heavy neutrino states, with new high mass states and extremely small mixing angles. However, there are theoretical models that decouple the mixing angles from the mass relations \cite{BUC}. This is the case if in the general mass matrix one imposes some internal symmetry that makes the matrix singular. Then the mixing parameters are bounded only by their phenomenological consequences. Another possible scenario for heavy neutrinos is in grand unified  extensions of the Standard Model such as $SO(10)$, $E_6$, as well as in mirror models. This new heavy neutrinos could be of the Dirac type. In this case we have no simple connection between mixing angles and mass ratios. 
\par
We are then lead to consider the possibility of new heavy neutrino states of Majorana and Dirac types, with  mixing angles with light neutrinos limited only by their phenomenological consequences. There are experimental bounds on heavy neutrino masses indicating that, if they exist, their masses must be greater than 80-100 GeV \cite{ZUB,PDG} and mixing angles between heavy and  light neutrinos are expected to be small. There is some model dependence on these results, but from radiative corrections \cite{PL} there is also no indication of new physics in this region. A recent work by Novikov \cite{NOV} suggests that in some models, new heavy neutrinos can have masses as low as 50 GeV. 
\par
The high precision measurements of the Z properties at LEP/SLC indicates that the mixing of the presently known fermions and possible new heavy states  small, of the order of $\sin^2 \theta_{{\mbox mix}}=10^{-2}-10^{-3}$. A recent estimate \cite{YPS} gives 
$\sin^2 \theta_{{\mbox mix}} < 0.0052$ with $95\%$ C.L. for the electron family. For the muon family we founded a stronger bound and for the tau family the bound must be weaker. This limit value is used throughout this paper for all curves and distributions and only the electron family is considered. Our results can also be extended to the other families with care in rescaling the mixing angles bound and background calculations.
\section{The Model}
\par
Single production of new heavy neutrinos in electron-positron colliders offers a clear possibility for a search in the neutrino mass region $M_N=(\sqrt{s}/2,\sqrt{s})$. We know experimentally that there are no new interactions in this kinematical region. So, after mixing the relevant part of the Lagrangian at LEP II energies is given by:
\begin{equation}
{\cal L}_{nc}=-\frac{g}{4c_W } {sin{\theta_{{mix}}}} Z_{\mu}\overline{\psi_N}\gamma^{\mu}
\left(1-\gamma_{5}\right)\psi_{\nu_e} + h.c..
\end{equation}
and
\begin{equation}
{\cal L}_{cc}=-\frac{g}{2\sqrt{2}} {sin{\theta_{{mix}}}}
W_{\mu}\overline{\psi_N}\gamma^{\mu}\left(1-\gamma_5\right)\psi_e+ h.c.
\end{equation}
where  N is the new heavy neutrino. For Dirac neutrinos we impose lepton number conservation and for Majorana neutrinos we must allow lepton number violation. There is some model (singlets, doublets, mirror neutrinos) dependence on the N-N-Z vertex but for the light-to-heavy neutrino vertex this dependence disappears \cite{AMY}.
\par
We are interested in the process $e^+e^- \longrightarrow \nu \, N $ and $ N \longrightarrow  e^\pm \, q_i\, \bar q_j $ since it gives a clear signature for heavy neutrinos and has a higher cross section than the pure leptonic final states and heavy neutrino pair production. As we will shown below in this paper, it also allows a clear separation between the signal and the Standard Model background. We have taken into account all first order contributions for this process.
\par 
In Fig. 1 we show the first order Feynmam diagrams that display the exchange of a single new heavy neutrino for the above process. In Fig. 2 we show the corresponding Standard Model diagrams that contribute for the same final state. We must also include in the Standard Model calculation, which is our background, all the diagrams of Fig. 1 with the heavy neutrinos replaced by light neutrinos. The single heavy Dirac neutrino for the electron family dominates the associated muon (and tau) family production since in the first case we have s and t channel exchanges, whereas in the last cases we have only s channel contribution. For single heavy Majorana one has to sum over final neutrino and anti-neutrino production and, in principle, to do the correct sum over the three lepton families. As we have the bound 
$\sin^2 \theta_{\mu} << \sin^2 \theta_{e}$, we have chosen not to sum over families, in order to reduce the number of new hypothesis and new parameters. Another point to be taken into account is the fact that the final state light neutrino is an experimentally undetected particle. So we must sum over all possible combinations whenever necessary.
\par
In the search for new particles a fundamental point to be clarified is the relation between the heavy neutrino  signal and the Standard Model background. This point was recently studied for future electron-positron colliders \cite{AMY} at $\sqrt s=$ 500 GeV and new electron-muon colliders \cite{PLB,CVE} where single  production of new heavy neutrinos was shown to be more important than pair production  \cite{ARS,DJO}. A similar study was also done for hadron-hadron colliders \cite{YPS,PAN}.
\par
In the present work we have done a detailed study of the process $e^+e^- \longrightarrow \nu \, e^\pm \, q_i\, \bar q_j $. Calculations for cross sections and distributions are straightforward, although rather lengthly. We have at our disposal efficient algebraic programs like CompHep \cite{HEP} that can perform this kind of calculations.  Hadronization of quarks was done with the well known program Phytia  \cite{PIT}. The complete hadronic reconstruction depends on each detector characteristics. In order to present our results as general as possible, we decided not to allow any hadronic decay after the hadronization process of quarks. The production of Majorana neutrinos in hadron colliders was recently \cite{PAN} done in the helicity amplitude formalism and found to give the same results of the CompHep package. These author found some  discrepancies with our previous results in ref.\cite{YPS}, mainly in the total width of the new heavy Majorana neutrino. This was due to the ghost contribution necessary to describe the W-boson longitudinal polarization. We have verified that the expression given in the appendix of ref.\cite{PAN} is the correct expression for the total width of the heavy neutrino. The important point for an experimental search is the comparison between  the signal for Majorana neutrinos with the Standard Model background. Since the Standard Model contribution is given by a large number of diagrams (see Figs. 1, 2),we have decided to employ the CompHep package for both the signal and background. After implementation of the Standard Model extentions here described, we have explicitly verified that our results are gauge independent. This procedure was  also confirmed by the authors of ref.\cite{PAN}, that verified that CompHep and  the helicity formalism for the amplitudes give the same results.   
\section{Results}  
\par
The total cross section for $e^+e^- \longrightarrow \nu \, e \, q_i\, \bar q_j $ at LEP II energy of $\sqrt {s}=$ 200 GeV is shown in Fig. 3. From this figure on we have summed over the final state charged leptons $ e^-$, $e^+$,  and employed  general detector cuts for final particles  $E_{e}> 5$ GeV and $-0.95 < cos \,\theta_{e} < 0.95$, where $\theta_{e}$ is the angle of the final charged lepton relative to the initial electron. The Standard Model background clearly dominates the signal. For both Dirac and Majorana production we have the same cross section if sums of final particles are only in the first family. We turn now our attention to  distributions and cuts that can improve the signal to background ratio. In order to make our calculations closer to experiment, we have hadronized all final state quarks using the Pythia program \cite{PIT}. All distributions are shown for the Majorana heavy neutrino. For the Dirac case we have a very similar pattern.
In Fig. 4 we display the invariant "charged lepton + neutrino" mass distribution $M_{e\nu}$. We note that the left scale applies to the Standard Model background and the right one applies to the signal curves for $M_N=$ 80, 100, 120 GeV. The background events are strongly concentrated at the W mass value. This suggests that we must include events whenever $M_{e\nu}< 75 $ GeV or $M_{e\nu} > 85$ GeV.
Another interesting variable is the total invariant visible "charged lepton + hadrons" mass distribution shown in Fig. 5 for $M_N=$120 GeV. The background has its maximum at 170 GeV and the signal is peaked at the heavy neutrino mass. An important point for this distribution comes from the fact that in the  models  that we are considering the heavy neutrino has a very narrow width. If this distribution is done with large bins, the signal is spread and lost. If the bin is narrow, the signal becomes more clear. Fig. 5 shows this effect for 1 GeV and 5 GeV bins.
The heavy neutrinos are very narrow resonances, in the  MeV range. The choice of bins as narrow as possible will be an experimental limitation of the available statistics. The LEP II results\cite{OPA} on the determination of the $W$ mass have already reached 1-GeV bins which will be sufficient for detecting heavy neutrinos in the 100-200 GeV mass region.
\par
Another point to be taken into account is the initial state radiation (ISR) and "beamstrahlung". Recent calculations \cite{JAD} for the Standard Model process $e^+e^- \longrightarrow $ 4-fermions at LEP II energies show that these contributions are at the few percent level. This is due to the fact that the 4-fermions process is dominated by real $WW$ production. The real $WW$ pair carries almost all the initial state energy, leaving practically no place for photon emission. This is also true for the case of new heavy neutrino real production at LEP II. We have explicitly checked for the mass region that we are considering, that there is little room  for ISR and "beamstrahlung" corrections. Therefore, no significant distortions occurs in the distributions presented in this paper but we decided to include ISR and "beamstrahlung" in all our results.
\par
The invariant hadronic mass is  peaked at the W mass for both the background and signal. We have then selected  the events with 70 GeV $ < M_{\mbox {hadrons}} < $ 90 GeV in order to improve the final state quark hadronization. Another useful cut comes from the angular correlation between  $\theta_e$, and the reconstructed final state hadronic angle relative to the initial electron, $\theta_{\mbox {hadrons}}$. In the next figures we have chosen the value $ ({\cos\theta_e}-1)^2 + (\cos\theta_{\mbox {hadrons}}+1)^2 > $ $(0.6)^2$. 
In Fig. 6 we show the invariant visible mass (charged leptons + hadrons) versus missing (neutrino) energy for background and signal (in arbitrary units) for $M_N=$100 GeV. In Fig. 6a we have done only the general detector cuts. For $M_N=$100 GeV the signal is already separated from the background but for higher masses this is no longer possible. The more general cuts discussed above can improve the signal to background ratio. Besides these cuts we have done in Fig. 6b the cut $ E_{\mbox{charged lepton}} < $ 40 GeV. The background is clearly below the signal. 
\par
In Table I we present a detailed analysis of the number of events expected for the signal and background. We have considered an integrated luminosity of 1000 pb$^{-1}$ and $M_{N}=$ 80, 100, 120, 150 GeV. This value for the luminosity can be rescaled for each of the LEP collaborations and for the present total value of 2500 pb$^{-1}$. In the last column we show the statistical significance for the signal number of events "S", relative to the background number of events "B". This is also shown in Fig. 7. We note that each point has different experimental cuts, besides the common detector cuts  $E_{e}> 5$ GeV and $-0.95 < cos\theta_{e} < 0.95$, and $M_{e\nu}< 75$ GeV or $M_{e\nu} > 85$ GeV. We can see  that masses up to 150 GeV can be attained with a very clear statistical significance. Higher masses could be investigated but with a lower experimental definition.
\par
 All the results presented here are for a new possible heavy Majorana neutrino using the $\sin_{\theta}^2$ upper limit. These results are the same for a Dirac neutrino, since we are considering only one family. 
\section{Conclusions}
The present work shows that the recent LEP II data can test the possibility of new Dirac and/or Majorana neutrinos with mass in the region $M_N=(\sqrt{s}/2,\sqrt{s})$ and mixing angles with light neutrinos in the range $10^{-2}-10^{-3}$. This was estimated for an integrated luminosity of 1000 pb$^{-1}$. The process  $e^+e^- \longrightarrow$ ``charged lepton + missing energy + hadrons" can give a clear signature for heavy neutrinos. For the models that we considered an important point is that the heavy neutrino width is very narrow and distributions must be done with narrow bins.
\par
\bigskip
{\it Acknowledgments:} This work was partially supported by the
following Brazilian agencies: CNPq, FUJB, FAPERJ and FINEP.

\LARGE
Figure Captions
\normalsize
\begin{enumerate}
\item Signal Feynman graphs for heavy Majorana  and Dirac  neutrino contribution to  $e^+e^- \longrightarrow \nu \, e \, q_i\, \bar q_j $.
\item Standard Model Feynman diagrams for $e^+e^- \longrightarrow \nu \, e \, q_i\, \bar q_j $.
\item Total cross section for the Standard Model background and for Dirac and Majorana heavy neutrinos at  $\sqrt{s}=$ 200 GeV and $\sin^2 \theta_{{mix}} = 0.0052$.
\item Invariant mass distributions for the system "charged lepton +neutrino". The left scale applies to the SM and the right scale applies to the signal.
\item Invariant mass distribution for the final state visible particles "charged lepton + hadrons", for $M_N=$ 120 GeV. The first figure is done with 5 GeV bins and the second one is done with 1 GeV bins. Both are in arbitrary units.
\item a. Invariant visible mass (charged leptons + hadrons) versus missing energy (neutrino) for background and signal for $M_N=$ 100 GeV (in arbitrary units). \par b. Same as Fig. 6a.  with the additional cuts as discussed in the text.
\item Statistical significance versus $M_N$ for different event selections  according to Table I.
\end{enumerate}
\end{multicols}

\begin{center}
\begin{tabular}{||c|c|c|c||} \hline
 & Signal & Background & $s /\sqrt B$ \\ \hline
M$_N$= 80 GeV & & & \\ \hline
Detector cuts & 148  &  2447 &  2.99 \\ \hline
Mass dependent cuts$^{\dagger}$ & 96   &  200 & 6.79  \\ \hline
Mass window $M_N \, \pm $ 10 GeV & 80   &  17 & 19.40   \\  \hline
Mass window $M_N \, \pm $ 5 GeV & 78 & 7 & 29.48 \\ \hline\hline

M$_N$= 100 GeV & & & \\ \hline
Detector cuts & 148  &  2447 &  2.99 \\ \hline
Mass dependent cuts$^{\dagger \dagger}$ & 103  &  148 & 8.47 \\ \hline
Mass window $M_N \, \pm $ 10 GeV & 102  &  80 & 11.40   \\  \hline
Mass window $M_N \, \pm $ 5 GeV & 101 & 43 & 15.40 \\ \hline\hline

M$_N$= 120 GeV & & & \\ \hline
Detector cuts & 100  &  2447 &  2.02 \\ \hline
Mass dependent cuts$^{\dagger\dagger\dagger}$ & 72  &  287 & 4.25 \\ \hline
Mass window $M_N \, \pm $ 10 GeV & 72  &  124 & 6.47  \\  \hline
Mass window $M_N \, \pm $ 5 GeV & 71 & 64 & 8.88 \\ \hline\hline

M$_N$= 150 GeV & & & \\ \hline
Detector cuts & 51  &  2447 &  1.03 \\ \hline
Mass dependent cuts$^{\dagger\dagger\dagger\dagger}$ & 15  &  57 & 1.99 \\ \hline
Mass window $M_N \, \pm $ 10 GeV & 15  &  17 & 3.64   \\  \hline
Mass window $M_N \, \pm $ 5 GeV & 15 & 9 & 5.00 \\ \hline
\end{tabular}
\end{center}
\par
$^{\dagger}$  $(\cos\theta_e-1)^2+ (\cos\theta_{\mbox{hadrons}}+1)^2 > (0.6)^2$ and $E_{e} < 40$ GeV.
\par
$^{\dagger\dagger}$ 70 GeV$ < M_{\mbox{hadrons}} < 90$ GeV,
$(\cos\theta_e-1)^2+ (\cos\theta_{\mbox{hadrons}}+1)^2 > (0.6)^2$ and $E_{e} < 40$ GeV.
\par
$^{\dagger\dagger\dagger}$  70 GeV$ < M_{\mbox{hadrons}} < 90$ GeV and
$(\cos\theta_e-1)^2+ (\cos\theta_{\mbox{hadrons}}+1)^2 > (0.6)^2$ and $E_{e} < 55$ GeV.
\par
$^{\dagger\dagger\dagger\dagger}$  70 GeV$ < M_{\mbox{hadrons}} < 90$ GeV, $\cos\theta_{\mbox{hadrons}} > 0.3$ 
and 40 GeV $< E_{e} < 70$ GeV.
\par
\bigskip
{\footnotesize Table I : Expected number of events  with $\cal L=$ 1000 $pb^{-1}$, for $M_{N}=80,100,120,150$ GeV.}

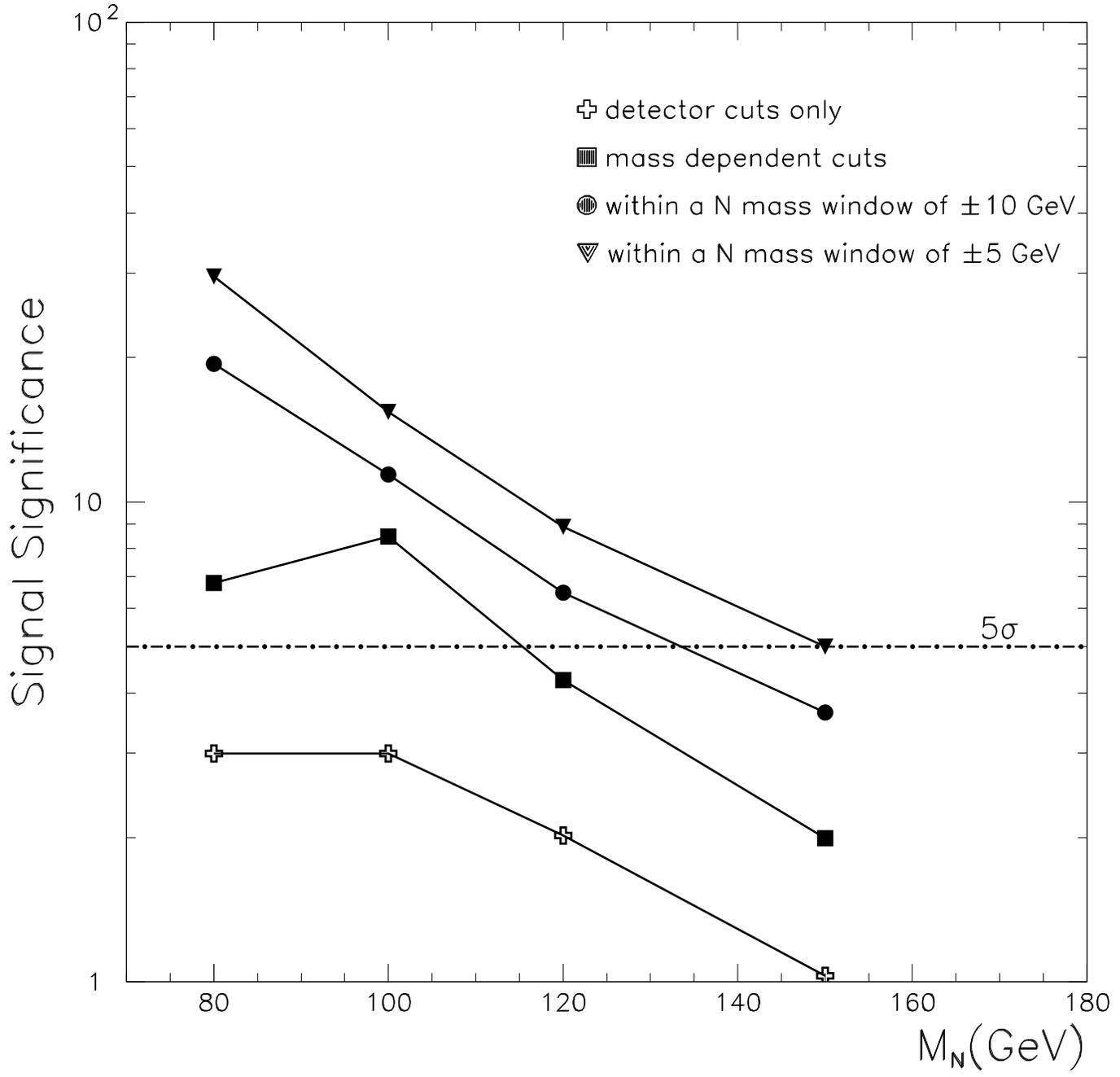
\end{document}